\pdfoutput=1

\documentclass[leqno,a4paper,12pt]{scrartcl}
\usepackage[centertags]{amsmath}
\usepackage{amsfonts}
\usepackage{amssymb}
\usepackage{multirow}
\usepackage{amsthm}
\usepackage{nicefrac}
\usepackage{pifont}
\usepackage[ansinew]{inputenc}
\usepackage{graphicx}
\usepackage[arrow, matrix, curve]{xy}

\begin{document}

\title{New ways of estimating excess mortality of chronic diseases: Insights from the illness-death model}
\author{Ralph Brinks\\Hiller Research Center for Rheumatology\\University Duesseldorf}
\date{}
\maketitle

\begin{abstract}
Recently, we have shown that the age-specific prevalence of a 
disease can be related to the transition rates in the illness-death model 
via a partial differential equation (PDE). In case of a chronic disease, 
we show that the PDE can be used to estimate excess mortality from prevalence 
and incidence. Applicability of the new method is demonstrated in a simulation and claims data
about diabetes in German men.
\end{abstract}

\section*{Introduction} 
Recently, we have shown that the age-specific prevalence of a 
health state or disease can be related to the transition rates in the illness-death model (IDM) 
via a partial differential equation (PDE) \cite{Bri14,Bri15c}. In case of a chronic disease, 
this relation can be used to estimate the incidence from a sequence of cross-sectional 
studies if information about mortality is available \cite{Bri15,Bri16}. 

In this paper, we demonstrate that it is also possible to estimate excess mortality from prevalence 
and incidence of a chronic disease, which can be useful for the analysis of data from disease 
registers or health insurance claims. For this, we examine the relations of the illness-death model and
associated PDEs. In this context, we derive a new PDE which generalises 
the PDE of Brunet and Struchiner \cite{Bru99}. In a simulation study, the new PDE is used 
to demonstrate how the excess mortality can be estimated directly. Furthermore, we present
an estimation method in the framework of Bayesian statistics. In an application of the Bayesian approach, 
we estimate the excess mortality of diabetes from claims data comprising 70 million Germans.

\section*{Illness-death model and associated partial differential equations} \label{sec:Gen} 
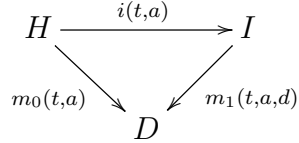
\begin{figure}[ht]
\centering
$$
\begin{xy}
\xymatrix{
H\ar[rr]^{i(t,a)}\ar[rd]_{m_0(t,a)} &   & I\ar[dl]^{m_1(t,a,d)} \\
                                    & D &
}
\end{xy}
$$
\caption{The transition rates $i, m_0, m_1$ between the
compartments $H$ (healthy), $I$ (ill), and $D$ (dead) in the illness-death model
depend on calendar time 
$t$ and age $a.$ The mortality rate $m_1$ additionally depends on the
duration $d$ of the disease.}
\label{fig:CompMod}
\end{figure}

We consider the illness-death model for chronic (i.e., irreversible) diseases 
shown in Figure \ref{fig:CompMod}. The considered population is split 
into the relevant disease states \emph{Healthy} ($H$) and \emph{Ill} ($I$). From either states
people can transit into the state \emph{Dead} ($D$). 
The transition rates between the three states are the incidence
rate ($i$), the mortality rate of the healthy ($m_0$) and the mortality rate 
of the diseased ($m_1$). These rates depend on the calendar time $t$ and 
on the age $a$. Additionally, the mortality rate $m_1$ depends 
on the duration $d$ of the disease. 

Let the numbers $H$ and $I$ denote the numbers of people in the 
respective states. To be more specific, $H(t, a)$ is the number of healthy
people aged $a, ~a \ge 0,$ at time $t;$ and $I(t, a, d)$ is the number of
diseased people aged $a, ~a \ge 0,$ at $t$ who are diseased for the duration 
$d, ~d \ge 0.$ We assume that the considered population is sufficiently large
that $H$ and $I$ can be considered as smooth functions. The total number of subjects 
aged $a$ at $t$ who have the chronic 
disease is $I^\star(t, a) = \int_0^a I(t, a, \delta) \, \mathrm{d}\delta.$  

\bigskip

Let us furthermore assume that the considered population is closed, i.e., 
there is no migration and that the disease is contracted after birth. 
The later assumption implies $I^\star(t, 0) = 0$ for all $t.$ Then, we can formulate 
following equations for the change rates of $H$ and $I$: 

\begin{eqnarray}
(\partial_t + \partial_a) \, H(t, a) &=& - \bigl ( m_0(t, a) + i(t, a) \bigr ) \, H(t, a) \label{e:PDE_H_ta} \\
(\partial_t + \partial_a + \partial_d) \, I(t,a,d) &=&  - m_1(t, a, d) \, I(t, a, d), \label{e:PDE_I_tad}
\end{eqnarray}

where $\partial_x$ means the partial derivative with respect to 
$x$, i.e., $\partial_x = \tfrac{\partial}{\partial x}$ for $x \in \{t, a, d\}.$

In addition to the PDEs \eqref{e:PDE_H_ta} -- \eqref{e:PDE_I_tad}), which describe the outflows from the 
states \emph{Healthy} and \emph{Ill}, we need the following initial conditions:
\begin{equation*}
\begin{split}
H(t - a, 0) & = H_0(t - a). \\
I(t,a, 0) & = i(t, a) \, H(t, a).
\end{split}
\end{equation*}
The first initial condition represents the number of (disease-free) newborns $H_0$, 
and the second initial condition describes the number of newly diseased persons, the incident cases.

\bigskip

In \cite{Bri15}, we have shown that the age-specific prevalence 
$p(t, a) = \tfrac{I^\star(t, a)}{I^\star(t, a) + H(t, a)}$ is the solution of a scalar PDE
that can be derived from the two-dimensional system \eqref{e:PDE_H_ta} -- \eqref{e:PDE_I_tad}. Here,
we choose a different approach. Instead of considering the age-specific prevalence $p(t, a)$, 
we follow the idea of Brunet and Struchiner and 
examine the \emph{prevalence-odds} $\pi(t, a) = \tfrac{I^\star(t, a)}{H(t, a)}$ \cite{Bru99}. 
Using the terminology $\partial = \partial_t + \partial_a,$ we obtain 
\begin{eqnarray*}
\partial \pi
    &=& \frac{\partial I^\star}{H} + \pi \, \left ( i + m_0 \right ) \\
    &=& \frac{-m_1^\star \, I^\star + i \, H}{H} + \pi \, \left ( i + m_0 \right ) \\
    &=& i + \pi \, \left ( i + m_0 - m_1^\star \right ).
\end{eqnarray*}

For the second equality we used $\partial I^\star = - m_1^\star \, I^\star + i \, H,$ which has been
proven in the Appendix of \cite{Bri15}. The rate $m_1^\star$ is defined as
\begin{equation}\label{e:defMstar}
m^\star_1(t, a) := 
\begin{cases}
\frac{\int\limits_0^a  m_1(t, a, \delta) \, I(t, a, \delta) \mathrm{d}\delta }
                               {\int\limits_0^a  I(t, a, \delta) \mathrm{d}\delta} & \textnormal{ for } I^\star(t, a) > 0\\
0 & \textnormal{ for } I^\star(t, a) = 0.
\end{cases}                       
\end{equation}

\paragraph{Remark:} The rate $m_1^\star$ may be accessible in epidemiological surveys
by choosing a sample population with representative distribution of disease duration.
However, in most practical cases, it is unknown because the distribution 
$\tfrac{I(t, a, d)}{\int_0^a  I(t, a, \delta) \mathrm{d}\delta}$ in Eq. \eqref{e:defMstar}
is not known.

\bigskip

Thus, we obtain following linear scalar PDE
\begin{equation}\label{e:dp}
\partial \pi = i - \pi \, \left ( i + m_0 - m_1^\star \right ),
\end{equation}

which shows how the temporal change of the prevalence-odds
$\partial \pi$ is governed by the rates in the illness-death model in Figure \ref{fig:CompMod}
and the value of the prevalence-odds itself.

\paragraph{Remark:} If $m_1$ does not depend on the duration $d$, we have $m_1^\star = m_1$
and Eq. \eqref{e:dp} is equivalent to the PDE (3) in Brunet and Struchiner \cite{Bru99}. Hence, Eq. \eqref{e:dp}
is a generalisation of the PDE of Brunet and Struchiner.

\paragraph{Remark:} Eq. \eqref{e:dp} is equivalent to 
\begin{equation}\label{e:PDE_p}
\partial p = (1-p) \, \bigl ( i - p \left (m_1^\star - m_0 \right ) \bigr ),
\end{equation}
which has been proven in \cite{Bri15}. With the definition $R = \tfrac{m_1^\star}{m_0}$
and letting $m$ be the overall mortality, $m = p \, m^\star_1 + (1 - p) \, m_0,$ 
then Eq. \eqref{e:PDE_p} becomes
\begin{equation}\label{e:PDE_pR}
\partial p = (1-p) \, \left ( i - m \; \frac{p \, (R - 1)}{1 + p \, (R - 1)} \right ).
\end{equation}

\bigskip

For our purpose of estimating the \emph{excess mortality} $\Delta m = m_1^\star - m_0$, 
Eq. \eqref{e:dp} is very useful, because it holds
\begin{equation}\label{e:excess}
\Delta m = \frac{i \, (1 + \pi) - \partial \pi}{\pi}.
\end{equation}

\bigskip

An advantage of the approach of Brunet and Struchiner lies in an explicit representation 
of the prevalence-odds in case the rates $i, m_0$ and $m_1$ are given. Then starting
from Eq. \eqref{e:PDE_H_ta} -- \eqref{e:PDE_I_tad} combined with the initial conditions of above, 
we obtain following equation by using calculus:

\begin{align}\label{e:p}
\pi(t, a) &= \int\limits_0^a i(t-\delta, a-\delta) \, \times \\
          & \qquad \exp \left ( - \int\limits_0^\delta \left \{ m_1(t - \delta + \tau, a - \delta + \tau, \tau) 
                                             - (i + m_0) \left (t - \delta + \tau, a - \delta + \tau \right ) \right \} \mathrm{d}\tau
              \right ) \mathrm{d}\delta. \nonumber
\end{align}

With $\pi = \tfrac{p}{1-p},$ we see that Eq. \eqref{e:p} is a generalisation of 
Eq. (1) in \cite{Bri15c}. One advantage of the explicit representation of 
$\pi$ in \eqref{e:p} is the possibility to (numerically) calculate $\pi$ with a 
prescribed accuracy, e.g. by Romberg integration \cite{Dah74}, which we will use 
in the examples below. Numerical solutions
of differential equations usually do not allow prescribed levels of accuracy. 

\bigskip

\section*{Examples and demonstration}
\subsection*{Direct estimation of excess mortality}
The first example is about a hypothetical chronic disease with all time-scales 
$t, a$ and $d$ playing a role. The incidence of the chronic disease is assumed to 
be $i(t, a) = \tfrac{(a-30)_+}{3000}$, which implies that the
disease affects only people aged 30 and older. 
The age-specific mortality rate of the non-diseased is 
chosen to be $m_0(t,a) = \exp(-10.7 + 0.1 a + t \ln(0.98)).$ 
In addition, we assume that the mortality $m_1$ of the diseased can be written as
a product of $m_0$ and a factor that depends only on the duration $d$:
$$m_1(t, a, d) = m_0(t, a) \times ( 0.04 (d - 5)^2 + 1).$$
Except for the time trend in $m_0,$ this example is the same as Simulation 2 in 
\cite{Bri14a}.

\begin{figure}[!ht]
\centerline{\includegraphics[keepaspectratio,width=4in]{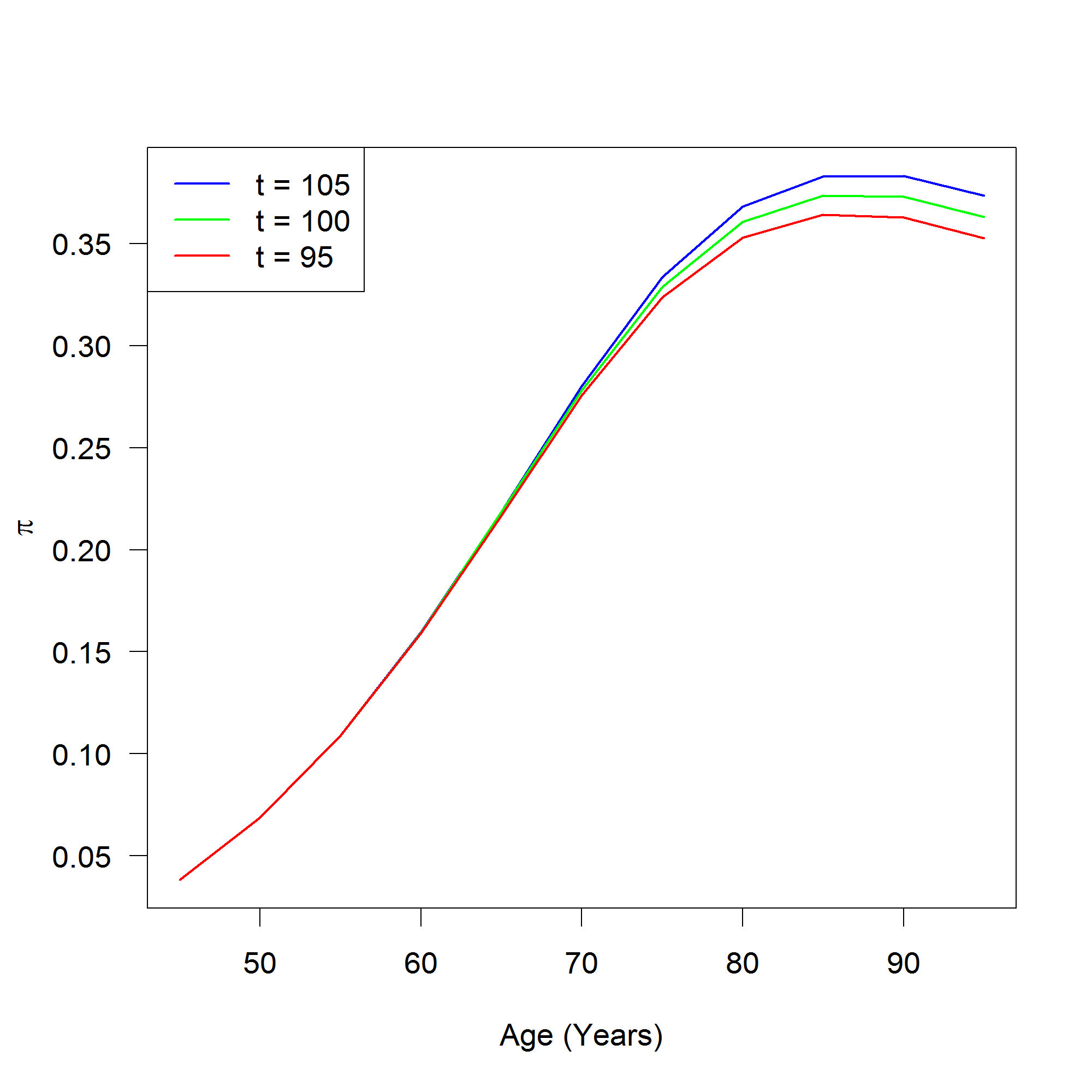}} 
\caption{Prevalence odds of the hypothetical chronic disease
in the years $t = 95$ (red), $t = 100$ (green), and $t =105$ (blue).} \label{fig:PrevOdds}
\end{figure}

For the example, we mimic the situation that we have three cross-sectional studies in the years
$t_0 = 95, t_1 = 100,$ and $t_2 = 105.$ We calculate the prevalence odds $\pi$ for these years via
Eq. \eqref{e:p}. Figure \ref{fig:PrevOdds} shows the prevalence-odds for the three years. 
Until age of about 70 years, the three prevalence-odds are virtually the same. For our example, we
additonally assume that we have the age-specific incidence rate available for the year $t_1 = 100.$
The aim is to estimate the excess mortality in $t_1$.

The proposed method to estimate the excess mortality $\Delta m$ in the year $t_1=100$ 
is direct application of Eq. \eqref{e:excess}. As assumed the incidence $i$ for $t_1 = 100$
is assumed to be given. The partial derivative $\partial \pi$ is approximated by 
following finite difference:
\begin{equation*}
    \partial \pi (t_1, a) \doteq \tfrac{1}{t_2-t_0} \; 
           \left [ \pi(t_2, a + \tfrac{t_2-t_0}{2}) - 
            \pi(t_0, a - \tfrac{t_2-t_0}{2}) \right ].
\end{equation*}

Then, we the excess mortality $\Delta m$ can be estimated by plugging these numbers into
Eq. \eqref{e:excess}. In case the mortality rate
$m_0$ of the non-diseased is known, $\Delta m$ is often expressed in terms of the hazard ratio 
$R = \tfrac{m_1^\star}{m_0}$ which can be obtained from
\begin{equation*}
R = 1 + \frac{\Delta m}{m_0}.
\end{equation*}

\begin{figure}[!ht]
\centerline{\includegraphics[keepaspectratio,width=4in]{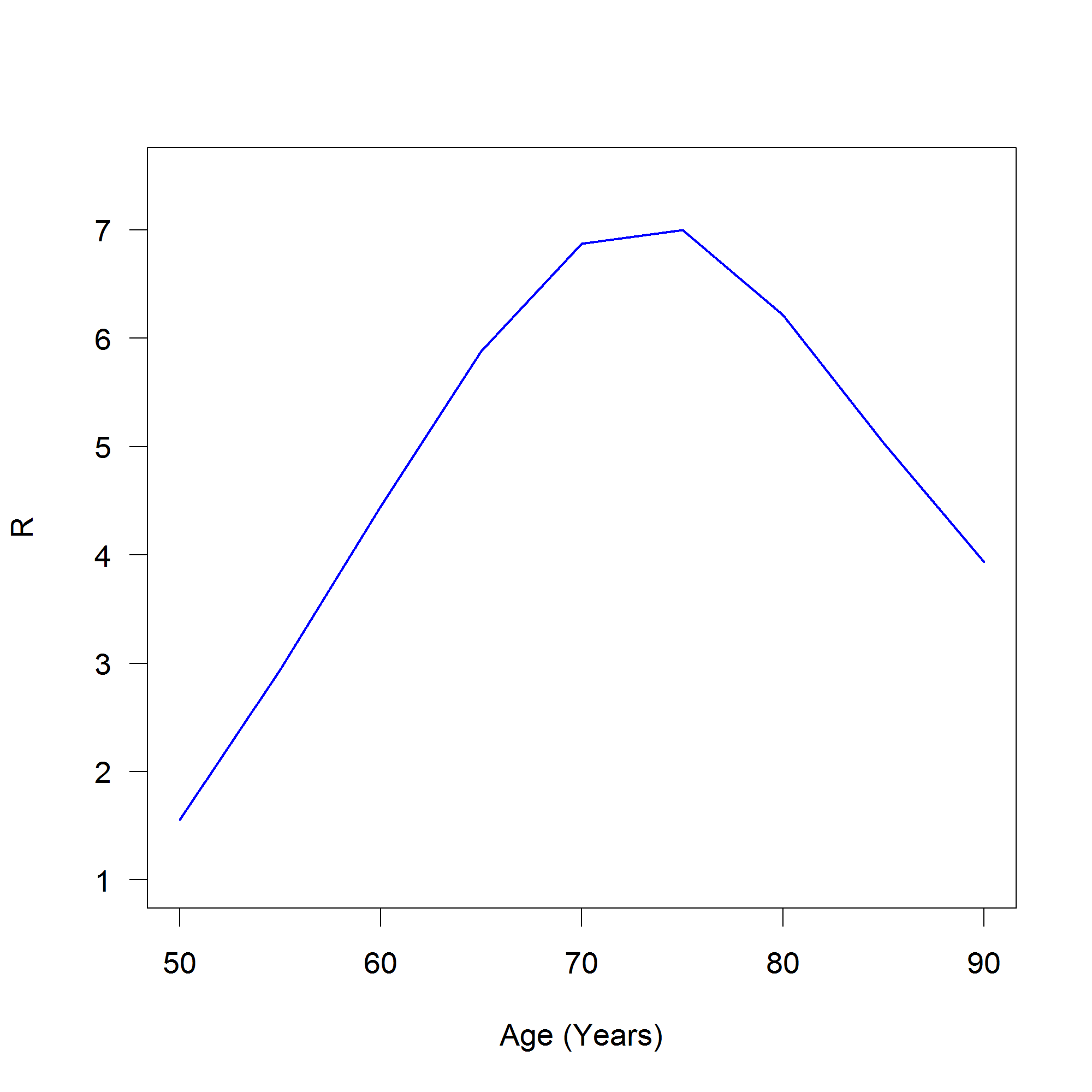}} 
\caption{Age-specific hazard ratio $R = \tfrac{m_1^\star}{m_0}$ in the year $t=100$.} \label{fig:HR}
\end{figure}

The age-specific HR expresses the mortality rate of the diseased people relative to the non-diseased
at the same age. For the hypothetical chronic disease we find the age-specific HR as in Figure \ref{fig:HR}.
The age-specific HR is peaking between age $a = 70$ and $a = 80$ and falling with increasing age.

In case, the mortality rate $m_0$ of the non-diseased population is not known, $\Delta m$ can also be
compared to the mortality rate $m$ of the general population. It holds 
$m = (1-p) \, m_0 +  p \, m_1^\star = \tfrac{1}{1+\pi} \, m_0 +  \tfrac{\pi}{1+\pi} \, m_1^\star$. 
Usually, the mortality of the general population is accessible from vital statistics of the federal 
statistical offices.

\subsection*{Bayes estimation of excess mortality}
\begin{figure}[!ht]
\centerline{\includegraphics[keepaspectratio,width=4in]{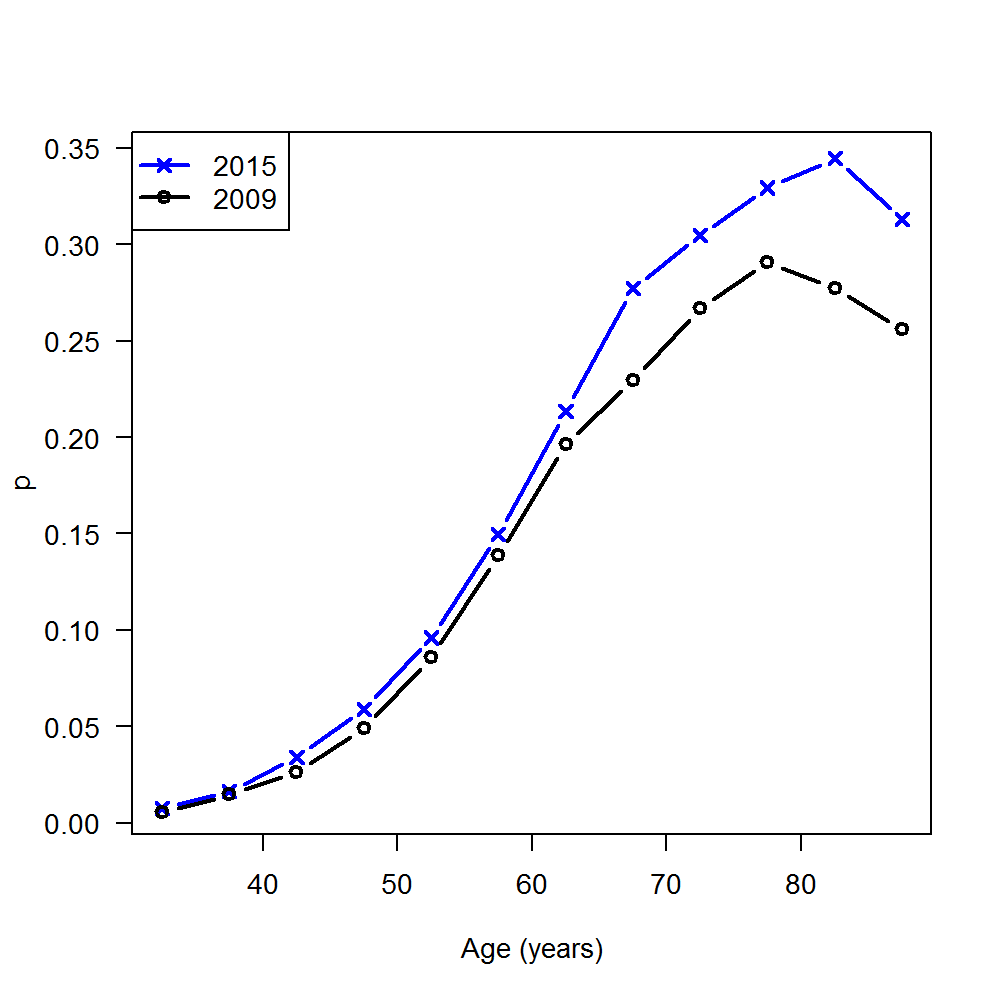}} 
\caption{Age-specific prevalence of diabetes in German men in 2009 (black) and 2015 (blue).} \label{fig:PrevDiabetes}
\end{figure}

The second example is about claims data from Germany during the years 2009 to 2015. Goffrier 
and colleagues reported the age-specific
prevalence of diabetes of German men in the years $t_0 = 2009$ and $t_1 = 2015$ as shown in Figure 
\ref{fig:PrevDiabetes}, \cite{Gof17}.

Based on the incidence rate ($i$) in 2012 (reported in Table 5 of \cite{Gof17}), our aim is to estimate
the age-specific hazard ratio $R(a)$ for the same year. For this, we use a Bayes approach. Motivated 
by empirical findings from the Danish diabetes register, we assume that the logarithm of the age-specific
HR $R$ approximately is a straight line in the age range 50 to 90 years of age (see Figure 5 in \cite{Car08}). 
Thus, we make the approach
\begin{equation*}
\ln R(a) = \ln R(50) + \frac{\ln R(90) - \ln R(50)}{90 - 50} \times a.
\end{equation*}

For $R(50)$ and $R(90)$ we use weakly informative prior distributions $R(50) \sim U(1.5, 4)$ 
and $R(90) \sim U(1, 2.5),$ where $U$ means the uniform distribution.
In Bayes terminology, our aim is to estimate a-posteriori distributions for 
$R(50)$ and $R(90).$ 

\begin{figure}[!ht]
\centerline{\includegraphics[keepaspectratio,width=4in]{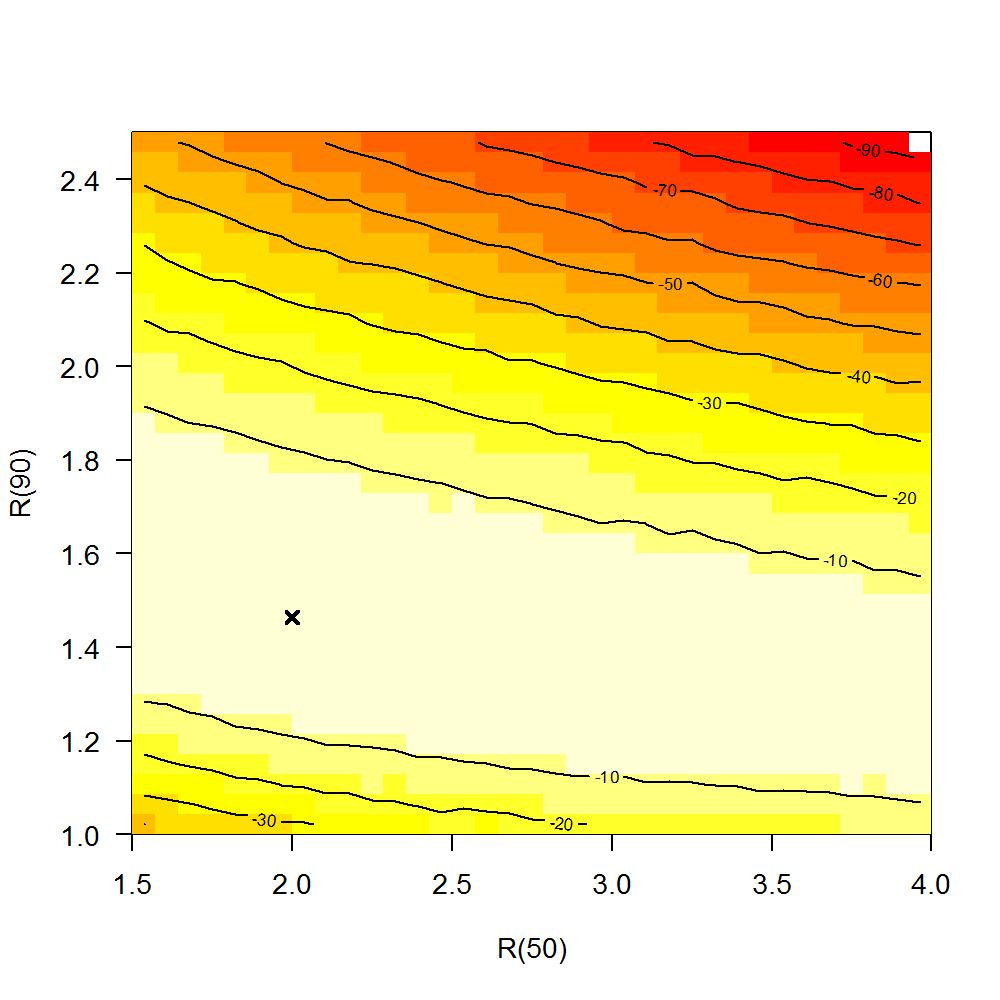}} 
\caption{A posteriori distribution of $R(50)$ and $R(90)$ with 
maximum a-posteriori estimator (black cross).} \label{fig:Aposteriori}
\end{figure}

This done by randomly drawing $R(50)$ and $R(90)$ from the prior distributions, solving
the PDE \eqref{e:PDE_pR} with initial condition $p(2009, a).$ For solving the PDE,
we use the Method of Characteristics \cite{Pol01} to convert Eq. \eqref{e:PDE_pR}
into an ordinary differential equation (ODE) and then solve the ODE by the Runge-Kutta Method
of fourth order \cite{Dah74}.
The calculated prevalence in 2015, $p(2015, a)$, is then compared with the 
observed prevalence in 2015 as shown as blue line in Figure \ref{fig:PrevDiabetes}.
Instead of the joint a-posteriori distribution, the log-likelihood of the deviation between observed
and calculated prevalence is computed. The results are shown in Figure \ref{fig:Aposteriori}. 
The black cross indicates the maximum a-posteriori 
(MAP) estimator for these data, which is given by $R_\text{MAP}(50) = 2.19$ and 
$R_\text{MAP}(90) = 1.46.$

\section*{Discussion} 
In this work, we have described how the illness-death model can be
used to obtain information about mortality in case prevalence and incidence are given. 
This allows insights into the excess mortality of people with chronic diseases compared
to the people without the disease or the general population. 

We presented two methods of estimating excess mortality, one direct estimation method and one
method in the context of Bayesian statistics. The methods can be used
if prevalence and incidence of chronic disease are given, e.g. in claims data \cite{Gof17} 
or in the setting of disease registers \cite{Sea04}.

\bibliography{denebBayes}

\bigskip

\emph{Contact:} \\
Ralph Brinks \\
Hiller Research Center for Rheumatology\\
University Clinics Duesseldorf\\
D-40225 Duesseldorf\\
\verb"ralph.brinks@ddz.uni-duesseldorf.de"
\end{document}